\documentclass[aps,prl,twocolumn,showpacs,floatfix,nofootinbib]{revtex4}
\usepackage{graphicx}
\usepackage{multirow}
\usepackage{amsmath,amsfonts}
\usepackage{times}
\newcommand{\ket}[1]{|#1\rangle}

\def\>{\rangle}
\def\<{\langle}

\begin{document}

\title{Efficient classical simulation of the semi-classical Quantum Fourier Transform }

\author{Daniel E. Browne}
\email{daniel.browne@merton.ox.ac.uk}
\address{Department of Materials and Department of Physics, University of Oxford, Parks Road, Oxford, OX1 3PU, UK}

\begin{abstract}
A number of  elegant approaches have been developed for the identification of quantum circuits  which can be efficiently simulated on a classical  computer. Recently, these methods have been employed to  demonstrate the classical simulability of the  quantum Fourier transform (QFT).
In this note, we show that one can demonstrate a number of simulability results for QFT circuits in a straightforward manner using Griffiths and Niu's semi-classical QFT construction [Phys. Rev. Lett. \textbf{76}, 3228 (1996)]. We then discuss the consequences of these results in the context of Shor's factorisation algorithm.

\end{abstract}


\maketitle

An important part of understanding the power of quantum computation relative to its classical counterpart is the identification of those circumstances where both classical and quantum computation have equal power. One important aspect of this is to identify quantum circuits which can be efficiently simulated on a classical computer, particularly when these circuits play a role in quantum algorithms believed to have an exponential advantage.

The standard description of a  quantum computation consists of the following steps. A register of quantum bits (qubits) is initialised in a certain state; Then a sequence of unitary quantum gates (the quantum circuit) are applied. Finally, each of the qubits is measured. The \emph{output} of a quantum computation is thus a classical bit string. This output  is not usually determinate -- rather bitstrings are returned by a quantum computation according to a particular probability distribution. We will say that a device which outputs bitstrings distributed according to this same probability distribution is \emph{simulating} the quantum computation.  Of particular interest are simulations on classical computers which are \emph{efficient}. This means that the resources needed to run the simulation (run-time, memory, etc.)  scale polynomially (or at the very least sub-exponentially) with the size of the problem, usually the number of qubits taking part in the quantum circuit. We shall call a family of quantum circuits acting on a given family of input states \emph{classically simulable} when an efficient method is known for its simulation on a classical computer.

Several classes of quantum circuit are known to be classically simulable when an appropriate simulation strategy is adopted. For example, classically simulable circuits include those consisting of  Clifford group gates acting on input qubits  in state $\ket{0}$ \cite{gottstabiliser}, circuits with restricted topological and depth properties \cite{markovshi,yoranshort,jozsa} acting on product state inputs and circuits where the entanglement is limited at every stage \cite{vidal,jozsalinden}.
 
These simulation strategies share a common feature -- the identification of a suitable classical data structure   to efficiently store the state or circuit, with efficient rules to update and evaluate it. In  the state vector formalism, where states are represented as vectors of complex amplitudes and unitary operators as square matrices, families of circuits over increasing numbers of qubits cannot generally be efficiently simulated, since the size of vectors and matrices increases exponentially. Nevertheless, Jozsa and Linden showed  \cite{jozsalinden} that certain classes of quantum circuit remain simulable in this framework, provided the entanglement in the states is restricted in a way that we shall describe below.
 
Recently, other data structures have been proposed. These include  the stabilizer formalism \cite{gottstabiliser}, the matrix product state description \cite{mps,vidal} and the weighted graph state \cite{WGS} approach, each  giving an efficient representation of multi-qubit states, and efficient tensor network descriptions \cite{markovshi,yoranshort,jozsa}  of quantum circuits. Each of these approaches has lead to new examples of  quantum circuits which are classically simulable.

Very recently, the classical simulability of the quantum Fourier transform has been investigated \cite{aharonovQFT,nadavQFT}. The quantum Fourier transform (QFT) \cite{shoralg} (over the field $\mathbb{Z}_{2n}$) is an  important family of quantum operations. It plays a special role in quantum computation theory, forming a key part of  Shor's factoring algorithm \cite{shoralg}.
The $n$-qubit QFT coherently transforms an input state $\ket{x}$ in the computational basis as follows;

\begin{equation}
\ket{x}\mapsto \frac{1}{\sqrt{n}}\sum_y e^{-\frac{i 2\pi}{n}x.y}\ket{y}
\end{equation}

where $\ket{x}$ represents a qubit-string $\ket{x_1\ldots x_n}$.

\begin{figure*}[ht]
\includegraphics[width=14cm]{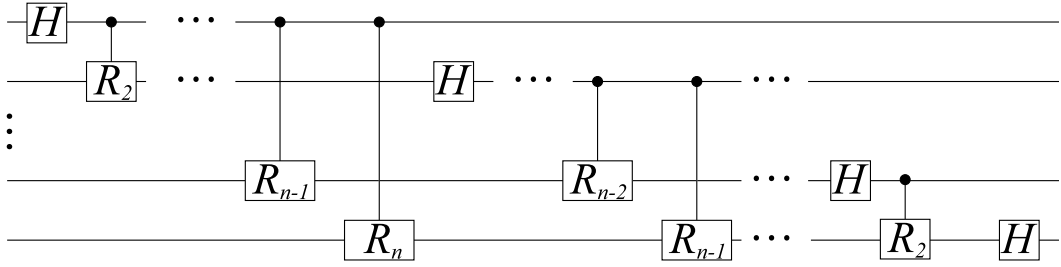}
	\caption{\label{exactqft}A schematic of the family of quantum circuits which implement the $N$-qubit QFT, as independently proposed by Deutsch and Coppersmith \cite{shoralg,ekertjozsa}. The controlled gates are controlled rotations $R_n=\exp[-i 2 \pi \sigma_Z 2^{-(n+1)})]$.  Note that this circuit also reverses the bit-order.}
\end{figure*}
\begin{figure*}[ht]
\includegraphics[width=14cm]{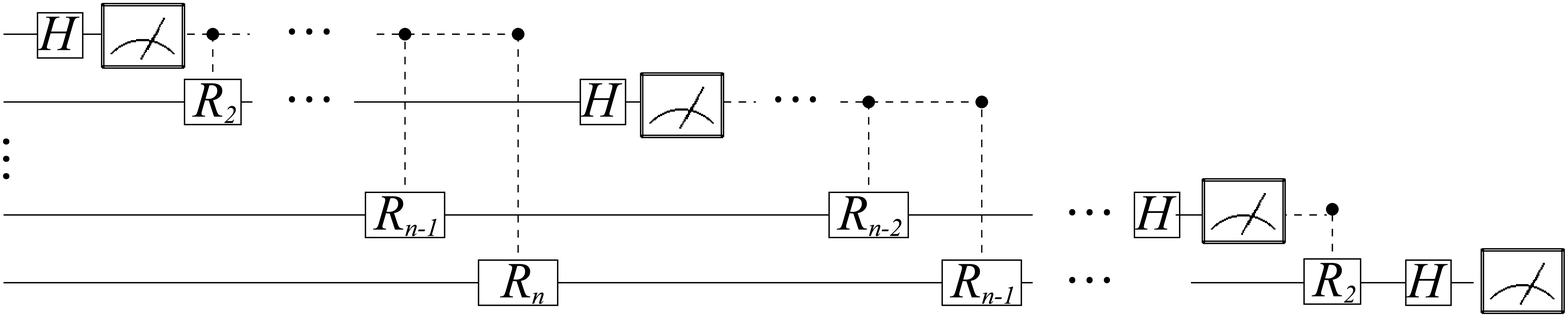}

	\caption{	\label{griffcirc}Griffiths and Niu's ``semi-classical'' circuit for the terminating QFT. The dashed lines represent the flow of classical control data.}
\end{figure*}
A family of  quantum circuits which efficiently realise this operation \cite{ekertjozsa} are illustrated  in figure \ref{exactqft}.
Other circuit implementations of the QFT have been proposed. Coppersmith  \cite{coppersmith} noted that the size of the controlled phase-rotations in figure \ref{exactqft} get progressively smaller to the point where their effect on the final state is insignificant. By neglecting these small rotations, one can produce an approximate QFT circuit with an output which has a high-fidelity compared to that of the exact circuit, but with a reduced number of gates. Cleve and Watrous \cite{clevewatrous} proposed an approximate QFT circuit which can  be parallelized to depth logarithmic in $n$.

We shall call a unitary \emph{terminating} if it occurs at the end a circuit, immediately preceding the (computational basis) measurement of all qubits.
  In Shor's algorithm, the QFT is terminating.  Griffiths and Niu \cite{griffithsniu} observed that when a controlled unitary is immediately followed by a measurement on the control qubit, a coherent  two-qubit implementation of the gate is not necessary. Instead, one can measure the control qubit before the target qubit, and use the measurement result as a classical control to determine whether to apply the single-qubit gate. This can be expressed via the following circuit identity, where the flow of a classical bit is represented by a dashed line:


\begin{equation}\label{niuidentity}
\includegraphics[scale=1]{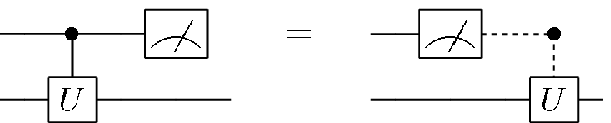}
\end{equation}

With this identity one can rewrite the circuit for  a terminating QFT so that it consists entirely of  measurements and adaptive single-qubit operations. This is illustrated in figure~\ref{griffcirc}.

Employing a tensor network description \cite{aharonovQFT,nadavQFT}, two groups of authors have recently presented independent proofs of the  classical simulability of the approximate QFT in a variety of contexts.
In \cite{aharonovQFT} a classical simulation method for the $n$-qubit Cleve and Watrous QFT circuit with sub-exponential $n^{O(\log n)}$ resource scaling. In \cite{nadavQFT} it was shown that any family of circuits with a constant number of Coppersmith's QFT circuit and additional logarithmic depth limited range circuits can be simulated with classical resources polynomial in $n$. In both \cite{aharonovQFT} and \cite{nadavQFT} the input state is assumed to be a  product state.

In this note, we shall show that adopting Niu and Griffiths' QFT circuit allows a very simple analysis of  the simulability properties of the terminating quantum Fourier transform.
We will show that a terminating QFT acting on product state inputs is classically simulable with a simple argument which takes just a few lines, in contrast to the elegant but somewhat complicated tensor methods employed in \cite{aharonovQFT} and \cite{nadavQFT}.
Furthermore, by combining this approach with  prior results we are able to extend the argument to cover a much wider class of non-product input states.

\textbf{\emph{Classical simulability of terminating QFT with product state inputs}}. 
 Niu and Griffiths' circuit for the terminating QFT  consists of a series of adaptive single-qubit unitaries and measurements. Since there are no entangling gates, the circuit can be thus simulated by following the evolution of individual qubits one by one. For each qubit this consists of a sequence of (adaptive) unitary gates followed by a measurement. Both the update of a single qubit state under the action of a single qubit gates and
 calculating the probabilities of the two measurement outcomes require a small (constant) number of classical calculation steps. By sampling from this probability distribution and using the generated classical bit as the  control for the next rounds of single-qubit gates a simulation of the complete circuit is achieved.  The full classical simulation consists of a linear number of simulated single qubit measurements and a quadratic number of single-qubit gates and is thus efficient.

\textbf{\emph{Classical simulability of terminating QFT with entangled inputs}}.
Jozsa and Linden showed \cite{jozsalinden} that entangled states across arbitrary numbers of qubits may be efficiently represented provided the state is \emph{$p$-blocked} - in other words, for some fixed $p$ the state can be written as a product of pure entangled states across subsets consisting of no more than $p$ qubits. Furthermore, they showed that any family of circuits is classically simulable, if the entanglement remains $p$-blocked at all stages of the computation.

Let us consider terminating QFT circuits with $p$-blocked input states. Since Niu and Griffiths' circuit contains no operations which can increase the entanglement of the state, the arguments of  \cite{jozsalinden} tell us that the circuit is classically simulable. 

We can also consider other kinds of entangled input states. There exist classes of states which do not not fulfill Jozsa and Linden's $p$-blocked criterion, but which still have an efficient classical representation. These include matrix product states (MPS) and graph states, which we shall consider in turn.

Matrix product states are states which can be written in the following manner \cite{mps,yoranshort,jozsa}.
\begin{equation}
\ket{\psi}=\sum_{i_1,\cdots,i_n=0}^1 C_{i_1 \cdots i_n}\ket{i_1}\cdots\ket{i_n}
\end{equation}
where the coefficients $C_{i_1 \cdots i_n}$ can be represented in matrix product form
\begin{equation}
C_{i_1 \cdots i_n}=\textrm{Tr}[M_1^{(i_1)}M_2^{(i_2)}\cdots M_i^{(i_n)}]
\end{equation}
where $M_1^{(i_1)},M_2^{(i_2)},\cdots, M_n^{(i_n)}$ are a set of matrices. The dimensions of each matrix is $D_{i}\times D_{i+1}$. The value $D_i$ represents the Schmidt number \cite{vidal} over the partition of the state between qubit $(i-1)$ and $(i)$. We label the maximal Schmidt number over all such partitions as $\chi$. The total number of parameters describing the state scales as $O(n \chi^2)$. If a family of states over increasing number of qubits $n$ is  bounded in $\chi$ then it can be efficiently represented in this framework.

Let us consider the action of the $n$-qubit terminating quantum Fourier transform on a set of input states with a matrix product state description with bounded $\chi$.  Vidal \cite{vidal} and Yoran and Short \cite{yoranshort} showed that to update a matrix product state description under the action of a single qubit unitary gate requires $O(\chi^2$) elementary operations \cite{vidal} and does not increase $\chi$. In addition, to update the MPS description after a computational basis measurement on one on the qubits \cite{yoranshort} requires $O(n \textrm{poly}(\chi))$ operations. Referring back to Niu and Griffiths' circuit we thus see that the action of the terminating QFT on a family of MPS input states with bounded $\chi$ is classically simulable.

Graph states \cite{graphstates} are a generalisation of the cluster states introduced by Briegel and Raussendorf \cite{briegelcluster}. Each state is associated with a graph consisting of vertices and edges connecting vertex pairs. Every vertex on the graph is associated with a qubit prepared in state $\ket{+}=(1/\sqrt{2})(\ket{0}+\ket{1})$. The graph state associated with the graph is generated by application of controlled-$\sigma_z$ gates between all each pair of qubits whose vertices are connected on the graph.

Single-qubit measurements on certain classes of graph states are universal for quantum computation \cite{1wqc}. However single qubit measurements on families of graph states with certain topologies  e.g. for graphs which are one-dimensional \cite{nielsen} or square lattice graph states with logarithmic width \cite{yoranshort,vandennestsim}  the evolution of the state under single-qubit measurements can be efficiently classically simulated. Since the terminating QFT can itself be implemented with adaptive single-qubit measurements, the same reasoning applies and with  all such classes of graph states as input, the terminating QFT will be classically simulable.

\textbf{\emph{Discussion}}.
Since the \emph{terminating} QFT is shown to be classically simulable for such a wide range of input states, one may ask whether  this can be exploited for the efficient simulation of Shor's algorithm.  In Shor's algorithm, states generated from a modular exponentiation circuit (applied to a uniform superposition of computational basis states) are fed into a terminating QFT.

The results above tell us that if the quantum states returned by the modular exponentiation belonged to any of the classes of states considered above, then in that case, the  factorisation algorithm  would be classically simulable. In \cite{jozsalinden} Jozsa and Linden consider the possibility of such states being $p$-blocked. They find that the modular exponentiation does, indeed, \emph{typically} generate non-$p$-blocked states, i.e. with entanglement ranging over an unbounded  number of qubits. It is interesting to note that their argument  does not prove that there are not special cases where $p$-blocked states do arise.

It is worth discussing what implications these results have for our understanding of where the non-classical power of Shor's algorithm lies. These results throw into relief the importance of the entanglement properties of the states generated by the first stage of the algorithm - the modular exponentiation.  One should not, however, rush to consider modular exponentiation alone to be the more important stage of the algorithm. This would be overhasty, not the least because modular exponentiation is also, when terminating, a classically simulable operation. This is easy to see, by  recalling that the input state to the modular exponentiation is  a uniform superposition of all $n$-bit string states. The output of the modular exponentiation is the state $\sum_x\ket{f(x)}$. A computational basis measurement on the output register will reveal one of the bit  strings $f(x)$, which will occur with probability $2^{-n}$. Thus an efficient classical simulation strategy is immediately apparent - one chooses a particular classical input string $x$ according to  the probability distribution of the input state (in this case a uniform distribution), and then one calculates $f(x)$ classically.
This emphasises that it is  the coherent interface between the modular exponentiation and the QFT which allows Shor's algorithm to exhibit its (believed)  speed-up.

\emph{\textbf{Conclusion.}}
We have shown that  Griffiths and Niu's semi-classical construction of the terminating quantum Fourier transform gives  a simple way to understand why the QFT, when considered in isolation, is classically simulable. Leveraging results from the literature, we have used the same approach to show that for a wide variety of classes of input state (with  efficient classical description) this  simulabilty property remains. The methods here could be applied to any terminating circuit which can be implemented via single-qubit measurement and adaptive single qubit unitaries. 
These results illustrate the special  status of operations occurring in the terminating position in a quantum circuit for questions of simulability and highlight that  component circuits that are individually classically simulable can still be combined  to form circuits with an apparent quantum speed up.
We hope that they also help shed a little more light on the fascinating question of from where quantum computation does and \emph{does not} get its power.

\acknowledgments

We would like to thank Richard Jozsa, Tony Short and Nadav Yoran for inspiring discussions and in particular to the latter two for giving early access to the results presented in \cite{nadavQFT}. 
This was supported by  funding from Merton College, Oxford and the EPSRC's QIPIRC.

\end{document}